\documentclass[fleqn,12pt]{wlscirep}

\usepackage[utf8]{inputenc}
\usepackage[T1]{fontenc}

\usepackage{lineno}
\usepackage{setspace}

\usepackage{caption}

\title{Making Sense of Nothing}
\author[1,*]{Douglas C. Leonard}
\affil[1]{Department of Astronomy, San Diego State University, San
  Diego, CA 92182-1221, USA}
\affil[*]{e-mail: dleonard@sdsu.edu}

\begin{abstract}
\end{abstract}

\begin{document}
\flushbottom
\maketitle
\thispagestyle{empty}
\doublespacing

\textbf{A model of the optical light detected following the merger of two neutron
  stars reveals polarization to be a unique probe of the geometry of the
  kilonova explosion that accompanied the gravitational waves. }
\vspace*{1 cm}

At precisely 12:41:06 UTC on 17 August 2017, a watershed moment in the history
of astronomy occurred when electromagnetic radiation -- photons of light --
were received from the same astronomical source which, 1.7 seconds earlier, had
triggered a gravitational wave detection\cite{Goldstein17}.  Dubbed AT 2017gfo
(or, alternatively, GW170817 or GRB 170817A), the gravitational and
electromagnetic messengers came courtesy of two stellar corpses -- neutron
stars -- spiraling together to generate a ``kilonova'', rippling the fabric of
space and generating electromagnetic fireworks spanning from radio waves to
gamma rays.  In one of astronomy's worst-kept secrets, news of the
unprecedented detections and ensuing festival of followup observations was
withheld until 16 October 2017, when an avalanche of papers simultaneously
appeared on the astronomy preprint archive 
(\url{http://arxiv.org/archive/astro-ph}), officially heralding the birth of
"multi-messenger" astronomy.  Writing in {\it Nature Astronomy}, Mattia Bulla
et al.\cite{Bulla19} take a first crack at extracting quantitative meaning from
a curious null result announced on that day: The optical light of the kilonova
was unpolarized\cite{Covino17}.

That it has taken a full year for a theoretical paper to plumb the expected
polarization properties of kilonova light reminds us that astronomical
polarimetry is a tricky business.  The observations are challenging, the data
reduction tedious and prone to false-positives, and the analysis and ultimate
conclusions frequently open to discordant interpretation.  Yet for all the
complexity, the potential windfall is enormous: Polarimetry can reveal the
geometry of unresolvable astronomical sources (that is, those that remain
``point-like'' in our sky, certainly true for AT 2017gfo at a distance of
roughly 130 million light years\cite{Hjorth17}).  Such information is
generally beyond the reach of any other technique.

Here is the essential idea.  The hot, expanding ejected material of just about
any type of stellar explosion (nova, supernova, kilonova, etc.)  contains an
abundance of free electrons at early times, which very effectively scatter --
and polarize -- light.  That is, the oscillating electric and magnetic fields
associated with each photon of the light, normally oriented in random
directions (unpolarized), become lined up (polarized) by the scattering
process.  Indeed, if we could spatially resolve such an object, we would expect
to measure changes in both the direction and strength of the polarization as a
function of position in the scattering ``atmosphere''.  For an unresolved
spherical source (or, more generally, a source that presents circular symmetry
in the plane of the sky), however, the directional components of the
polarization cancel exactly, and yield zero net polarization (Fig. 1a).  If,
however, the source has regions of its electron-scattering atmosphere blocked
by obscuring material, incomplete cancellation occurs, and a net polarization
results (Fig. 1b).  Determining how much polarization is to be expected is
the job of the theoretical modeler, and will in general be a strong function of
the assumed initial physical conditions, how these conditions evolve with time,
and the viewing orientation to the system.

In the first polarimetry data ever reported for a kilonova, Covino et
al.\cite{Covino17} present five measurements of AT 2017gfo sampling from 1.46 to
9.48 days after the neutron-star merger.  All epochs yielded no intrinsic
polarization to varying degrees of statistical significance.  Naively, such
null results would seem to admit two possibilities: (1) An unobscured view of a
circularly-symmetric electron-scattering atmosphere (Fig. 1a); or (2) an
atmosphere in which the photons are not being predominantly scattered by --
and, hence, polarized by -- free electrons.  To sort out which is at work,
Bulla et al. undertake the first complete computational modeling of the
polarization expected for early-time kilonova light.

They begin by adopting the most popular model for coalescing neutron stars,
which posits that a small amount (typically a few hundredths of a solar mass)
of radioactive, neutron-rich nuclei gets ejected from the system at high
velocities\cite{Li98}.  These radioactive nuclei subsequently decay,
synthesizing heavy elements and powering the electromagnetic transient (the
kilonova).  They further tweak this basic model by segregating the ejected
material into two distinct components.  First, an equatorial region of
extremely high line opacity (photons are absorbed and reemitted many, many
times) produced by electronic transitions in atoms of heavy elements ---
specifically, lanthanides, which have atomic numbers ranging from 57 to 71.
Second, a much lower-opacity polar region of ejecta that consists of
``lanthanide-free'' atoms.  Such a two-component system is currently very
popular, and thought by many (but, not all) to be demanded by both
theory\cite{Kasen15} and observation\cite{Pian17}.  Demonstrating unambiguously
that the proposed lanthanide-free region exists would be a major advance.

\begin{figure}[ht!]
 \vspace*{-0.5 cm}
\begin{center}
 \hspace*{0.5 cm}
\rotatebox{270}{
 \scalebox{1.4}{
 \includegraphics[width=3.4in]{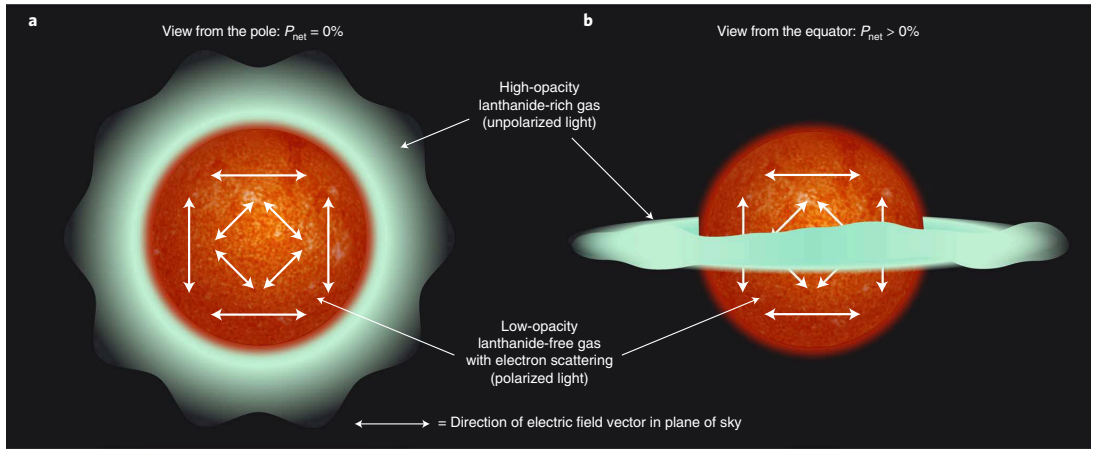} }}
\captionsetup{labelformat=empty}
\vspace*{-5 cm}
 \caption{ {\bf Fig. 1 | The polarization of light expected from a kilonova 1.5
     days after the neutron star merger.}  Significant electron-scattering
   opacity --- producing highly polarized photons --- exists in the
   lanthanide-free region of material ejected in the polar directions, whereas
   line-scattering opacity -- producing unpolarized photons --- dominates in
   the lanthanide-rich region of material ejected in the equatorial plane.
   {\bf a,} When viewed from above, an unresolved source yields zero net
   polarization ($P_{\rm net} = 0\%$) due to complete cancellation of the
   directional components.  {\bf b,} When viewed from the equator, however, the
   partial obscuration of the electron-scattering region results in incomplete
   cancellation, yielding a non-zero net polarization ($P_{\rm net} > 0\%$).  }
   \label{plot1}
\end{center}
\end{figure}

Of particular significance to polarization studies is the fact that any
``bound-bound'' line scattering experienced by a photon completely destroys its
polarization.  Thus, we may anticipate that the resulting net polarization for
a kilonova will depend strongly on the relative importance and locations of
electron scattering (highly polarizing) compared with line scattering
(depolarizing) in the atmosphere.

Armed with this physical model, Bulla et al. produce the polarization
characteristics of the emerging optical kilonova light as functions of both
time and viewing angle.  The headline?  We must catch kilonovae early!  Beyond
~two days post-merger, all polarization vanishes, rendering polarimetry
impotent as a geometric probe.  This happens due to the overwhelming dominance
of line opacity compared with electron-scattering in all regions of the ejected
material (both equatorial and polar) once it has expanded and cooled
sufficiently.  In this regard, the null polarization observed for AT 2017gfo
during the last four epochs (all > 2 days post-merger) is both expected and
unremarkable.

But at earlier times, the situation is profoundly different.  Large
polarizations are, indeed, possible, and are highly dependent on viewing angle:
Equatorial views at 1.5 days produce expected polarization levels approaching
$1\%$ while polar views yield nothing.  The reason is geometry.  Even at these
early times, polarization from electron-scattering can only be generated by the
lanthanide-free polar regions; in the equatorial regions, line-scattering by
the lanthanide elements is found to always completely depolarize the light.
Thus, we have a special --- and, fleeting --- situation in which an obscuring
equatorial ``belt'' partially blocks an electron-scattering atmosphere.  This
generates a net polarization for equatorial views (Fig. 1b) but, since there is
no obscuration, zero polarization for polar views (Fig. 1a).  Intermediate
orientations fall in between.  (Note that the shape of the electron-scattering
photosphere is shown to have surprisingly little impact on the resulting
polarization, and so it is assumed spherical.)  The upshot of this modeling is
that the null polarization measurement made 1.46 days post-merger --- formally,
the polarization is measured to be less than $0.18\% {\rm\ at\ } 95\%$
confidence --- constrains the viewing orientation of AT 2017gfo to be within
$\sim 65$ degrees of being pole-on.

These new models beg for more data from future events, since they are at once
eminently falsifiable, potentially restrictive, and alluringly predictive.
Falsifiable since any detection of intrinsic polarization beyond a few days
post-merger would demand a major rethink of a popular model's assumptions.
Potentially restrictive since unanimously null polarization measured for many
objects observed at sufficiently early times (and, presumably, viewing
orientations) would severely limit the possible vertical extent of the proposed
lanthanide-rich region in kilonovae.

Of course, a definitive detection of early-time polarization that rapidly falls
to zero would be hailed as soaring validation of the model's basic
components. But would it, as Bulla et al. assert, ``unambiguously reveal the
presence of a lanthanide-free ejecta component''?  To some, perhaps.  But
polarimetry is a tricky business.

\newcommand{\apj}{Astrophys. J.}
\newcommand{\apjl}{Astrophys. J. Lett.}      
\newcommand{\mnras}{Mon. Not. R. Astron. Soc.}   
\newcommand{\nat}{Nature}

%\bibliography{/Users/leonard/misc/all_refs}

\end{document}